\documentclass[%
 reprint,
superscriptaddress,
footinbib,
 amsmath,amssymb,showpacs,
 aps,
prl,
]{revtex4-1}

\usepackage{graphicx}
\usepackage{dcolumn}
\usepackage{bm}
\usepackage{subfig}
\usepackage{xfrac}
\usepackage{amsmath}
\usepackage{hyperref}
\usepackage{multirow}

\usepackage{color}

\DeclareCaptionJustification{justified}{\leftskip=0pt \rightskip=0pt \parfillskip=0pt plus 1fil}
\newcommand{\beqar}{\begin{eqnarray}}
\newcommand{\eeqar}{\end{eqnarray}}

\newcommand{\bcen}{\begin{center}}
\newcommand{\ecen}{\end{center}}

\newcommand{\ra}{\rangle}
\newcommand{\la}{\langle}

\newcommand{\beq}{\begin{equation}}
\newcommand{\eeq}{\end{equation}}
\newcommand{\beqa}{\begin{eqnarray}}
\newcommand{\eeqa}{\end{eqnarray}}

\begin{document}

\title{Time-dependent harmonic potentials for momentum or position scaling}
\author{J. G. Muga}
\affiliation{Department of Physical Chemistry, University of the Basque Country UPV/EHU, Apdo 644, 48080 Bilbao, Spain}
\author{S. Mart\'\i nez-Garaot}
\affiliation{Department of Physical Chemistry, University of the Basque Country UPV/EHU, Apdo 644, 48080 Bilbao, Spain}
\author{M. Pons}
\affiliation{Department of Applied Physics I, University of the Basque Country, UPV/EHU, 48013 Bilbao, Spain}
\author{M. Palmero}
\affiliation{Department of Applied Physics I, University of the Basque Country, UPV/EHU, 48013 Bilbao, Spain}
\author{A. Tobalina}
\affiliation{Department of Physical Chemistry, University of the Basque Country UPV/EHU, Apdo 644, 48080 Bilbao, Spain}
\begin{abstract}
Cooling methods and particle slowers as well as  accelerators are basic tools for fundamental research and applications in different fields  
and systems. We put forward a generic mechanism to scale  the momentum of a  particle, regardless of its initial position and momentum, by means of a  transient harmonic potential. The design of the time-dependent frequency makes use of a linear invariant and inverse techniques drawn from ``shortcuts to adiabaticity''.  The timing of the process may be decided beforehand and its influence on the system evolution and final features is analyzed. We address quantum systems but the protocols found are  also valid for classical particles. Similar processes are possible as well for position scaling. 
\end{abstract}

\maketitle

{\it Introduction.} 
Particle slowers and accelerators are basic tools for fundamental research and applications in different fields covering a huge range of systems, from high energy physics to atomic and molecular physics. Zeeman \cite{Phillips1982} or Stark slowers \cite{Bethlem1999}, 
optical slowers \cite{Fulton2004}, magnetic inverse coil-guns \cite{Narevicius2007,Dulitz2014}, and delta-kick cooling (DKC) \cite{Chu1986}, for example, have played a central role to develop cold and 
ultracold physics, while accelerators are needed to launch beams for controlled collisions, deposition \cite{Yagi1977}, or implantation \cite{Hamm2012}  at chosen  speeds. For such a vast domain of systems and conditions many different techniques have been developed. 
A broad family of methods applies  electromagnetic fields  adapted to the particle type and the operation, taking into account  if the particle is charged,  its magnetic moment, its dipole moment, its polarizability, or  
if it allows for cyclic  transitions. The results often depend heavily on the initial states, initial location, velocity, or spreads, and 
methods that could suppress or mitigate these dependences are of general interest.       

In this paper we find a simple, generic mechanism, and work out protocols, to scale the momentum of a classical particle, or of a quantum wave packet. The scaling can speed up or slow down the particle by a predetermined factor; this factor  could even be negative, to produce a ``momentum mirror''.  The main features of this mechanism are system independent, the only formal requirement is that the particle is subjected to a  transient harmonic potential with time-dependent frequency during a prearranged  duration. 
The specific system will determine  the practical details on how the  
harmonic potential is implemented, using optical, magnetic, electrical or mechanical means. 
An astonishing property of the  protocols described below is that  
the scale factor is the same for all initial conditions, i.e. for arbitrary  quantum wave packets or for all initial positions and momenta of the 
classical particles. While, in principle, information on the exact initial condition is not needed to perform the scaling, 
practical considerations will of course
set limits. These limits are not fundamental, but depend on the spatial, energetic, and temporal domain in which the needed harmonic potential can  be effectively implemented in some specific setting.  

The theory behind the time-dependent protocols for the harmonic potential makes use of an invariant of motion linear in position and momentum.
Basically, we deal with an inverse problem, where the Hamiltonian is found from the 
desired dynamics encoded in the invariant, along the lines of the set of inverse techniques known as ``shortcuts to adiabaticity'' \cite{Torrontegui2013,Guery2019}. 
The theory is worked out here for a quantum particle represented by a wave packet but the resulting protocols apply equally well to classical particles since, as it is well known, harmonic potentials lead to classical equations of motion for the expectation values of position and momentum. In fact the dynamics of an arbitrary wave packet  can be exactly reproduced by swarms of classical particles using the Wigner representation to fix the, possibly negative, ``weighting factors'' \cite{Muga1993}.       

We shall present first the theory and deduce the protocols. 
Then we provide expressions for the time dependence of expectation values of position and momentum   
for a chosen scale factor, as well as expressions for   
second order moments for positions and momenta in terms of the initial values. This is valuable information
to set both practical limits and design details depending on the intended target and resources available. 
We end the paper by considering related processes, in particular the scaling of positions, i.e.,  focusing or antifocusing.

{\it Lewis-Riesenfeld invariants.}   
Lewis-Riesenfeld   ``time-dependent invariants'' are operators whose expectation values 
remain constant for states driven by the 
associated  time-dependent Hamiltonian \cite{Lewis1969}. The time-dependent eigenvectors of the invariant 
conform a 
convenient basis, since their probabilities remain constant along the evolution. The  phases can be chosen to 
make each eigenvector a  solution of the time-dependent Schr\"odinger equation. 
This structure has been used systematically to inverse engineer  Hamiltonians from desired faster-than-adiabatic dynamics since  \cite{Chen2010_063002},  
for operations to control internal or motional states. 
Specifically in harmonic systems, most applications have made use of quadratic invariants in positions and momenta. The existence of linear invariants was known \cite{Castanos1994,Guasti2003,Lohe2009} but has not been 
exploited for inverse engineering. The bias towards quadratic invariants in most inverse engineering applications is in part explained 
by the fact that an ``Ermakov'' quadratic invariant may be set to commute with the harmonic oscillator Hamiltonian at initial and final process times \cite{Chen2010_063002}. Thus  fast  expansions, transport, rotations, or splittings between initial and final traps can be designed  so that  the final energy is the same as if the process had been very slow, i.e., adiabatic \cite{Guery2019}.  Instead, the linear invariants offer the possibility to control (scale) other observables, such as the momentum, the position, and therefore kinetic or potential energies. The linear invariant eigenvectors provide  continuum representations well adapted to 
processes  where  the initial and final harmonic frequencies 
vanish, a challenging limit for the discrete representations associated with the conventional Ermakov invariant. 

{\it Linear invariant.}
The Hamiltonian of a particle subjected to a harmonic potential with its center fixed 
at the origin and time-varying frequency is given by  
\beq
    H(t) = \frac{ {p}^2}{2m }
+ \frac{ m}{ 2}  \omega^2(t)  {q}^2. 
    \label{hamiltonianmw}
\eeq
Here we consider $q$ and $p$ noncommuting operators, but the same symbols may represent $c$-numbers 
in wave function representations, or as conjugate variables of a classical particle. The context should avoid any confusion. 
The linear combination of operators (dots
stand for time derivatives hereafter) \cite{Castanos1994,Guasti2002,Guasti2003}
\begin{equation}
     G(t)=u(t)  p - m\dot u(t)  q,
     \label{invaG}
\end{equation}
satisfies the invariant equation $i \hbar \partial  G / \partial t - [ H, G] = 0$, 
provided the reference trajectory $u$  satisfies   
\beq
\ddot{u}+\omega^2(t) u = 0,
\label{dyneqs}
\eeq
{which is a  classical equation of motion  for a particle driven by a ``classical'' Hamiltonian (\ref{hamiltonianmw}).  
For any quantum state evolving with $H(t)$, the expectation value of $G(t)$ at time $t$ is the  Wronskian $W(t)=W[u(t),\langle q\rangle(t)]$ times $m$, where both functions in the argument evolve
classically, i.e. following  a harmonic oscillator equation (\ref{dyneqs}), according to Ehrenfest's theorem. $\la G\ra$ is indeed invariant   
as $\dot{W}(t)=0$  
using Eq. (\ref{dyneqs}).
This result does not depend on the particular state 
so 
the expectation values can be substituted by operators in Eq. (\ref{invaG}). 
Here we shall consider only real solutions $u$. 

A corresponding quadratic invariant takes the form 
\beq
 I=\frac 1 {2m}  G^{\dagger} G= \frac{u^2  p^2}{2m} + \frac{m}{ 2} \dot u^2  q^2 
 -  \frac{1}{2}u\dot{u}(p  q + q  p).   
\label{inva}
\eeq
(To get the Ermakov quadratic invariant \cite{Lewis1969,Guery2019} $u$ has to be made complex,
see e.g.  \cite{Guasti2002}.) 
By imposing the boundary conditions at initial and final times $t_b=0, t_f$, 
\beq
\omega(t_b)=0,\;\;
\dot{u}(t_b)=0,
\label{bc1D}
\eeq
which also imply $\ddot{u}(t_b)=0$, see Eq. (\ref{dyneqs}), we find 
$G(t_b)=u(t_b)p$,  proportional to the  momentum. Thus, the final and initial  momenta are 
proportional to each other for any wave packet, $\langle p\rangle_f=({u_0}/{u_f})\langle p\rangle_0$, with a corresponding relation for  kinetic energies
due to the associated quadratic invariant,
%
$
E_f=(u_0/u_f)^2E_0,
$
%
where we use shorthand notations $u_0=u(0),\; u_f=u(t_f)$, and generally 
subscripts $f$ and $0$ for final and initial times.  The scaling does not only affect expectation values but also each
momentum component as we shall see. 
To design a harmonic slower 
or accelerator we first choose the scaling factor ${u_0}/{u_f}$, and a $u(t)$ that  satisfies the  
boundary conditions (\ref{bc1D}) and the scaling factor. $\omega(t)$  is found from Eq.   (\ref{dyneqs}) as 
\beq
\omega^2(t)=-\ddot{u}(t)/u(t).
\eeq 
With the chosen boundary conditions the eigenvectors of $G(t_b)$ or $I(t_b)$  are 
 plane waves, i.e., not square integrable, but they form a  valid and useful basis. The (constant-in-time) eigenvalues of $G(t)$ can be conveniently computed at time $0$ as $\lambda=u_0p_0$.  The initial plane wave momentum $p_0$ will play the role of integration variable to expand the wave functions. 
At an arbitrary time the eigenvectors of $G(t)$, $G(t)|\phi_{p_0}(t)\ra=u_0p_0|\phi_{p_0}(t)\ra$, may be calculated as
\beq
\phi_{p_0}(q,t)=\frac{e^{i\varphi_{p_0}(t)}}{h^{1/2}}e^{{i}(u_0p_0 q+m\dot{u}_tq^2/2)/(\hbar u_t)}.
\label{eigen}
\eeq
The phase $\varphi_{p_0}(t)$ is chosen so that Eq. (\ref{eigen}) represents a solution of the time-dependent Schr\"odinger equation,
and it is found by inserting Eq. (\ref{eigen}) into the Schr\"odinger equation, 
\beq
e^{i\varphi_{p_0}(t)}=\left(\!\frac{u_0}{u_t}\!\right)^{\!\!1/2}\!e^{-i\frac{p_0^2}{2m\hbar} {\cal I}_t},
\label{phase}
\eeq
{where}  ${\cal I}_t=\int_0^t {dt' u_0^2}/{u_{t'}^2}$,
%
and in general we use the subscript $t$ as a shorthand for the argument $(t)$.
The 
factor $h^{-1/2}$ in the eigenvector (\ref{eigen}) is chosen to have delta-normalized momentum plane waves at time $t=0$, 
$\la q|\phi_{p_0}(0)\ra=\la q|p_0\ra$, i. e., $\la p_0|p_0'\ra=\delta(p_0-p_0')$. 
Instead, at final time, $\la x|\phi_{p_0}(t_f)\ra=e^{i\varphi_{p_0}(t_f)}\la q|p_0 u_0/u_f\ra$. The invariant eigenstate 
that starts as a plane wave with momentum $p_0$ ends being  
proportional to a plane wave with momentum $p_f=p_0 u_0/u_f$. 
An arbitrary wave function may be expanded in the basis of functions (\ref{eigen}) as
\beqa
\psi(q,t)&=&\left(\frac{u_0}{hu_t}\right)^{\!\!1/2}\!
\int  \!dp_0 \exp\left[{\frac{i}{\hbar u_t}(u_0p_0 q+m\dot{u}q^2/2)}\right]
\nonumber\\
&\times&\exp\left({-i\frac{p_0^2}{2m\hbar} {\cal I}_t }\right) \la p_0|\psi(0)\ra.
\label{wave}
\eeqa
Integrating first over $q$ in the (implicit) triple integral $\int dq |\psi(q,t)|^2$ gives a delta function in momentum so 
$\int dq |\psi(q,t)|^2=\int dp_0 |\la p_0|\psi(0)\ra|^2$, i.e., the norm is conserved at all times.  

Here we choose polynomial trajectories
for simplicity, with the coefficients fixed so that $\dot u(t_b)=\ddot u(t_b)=0$, $u(0)=u_0,\; u(t_f)=u_f$, 
\beq
u(t)=u_0+(u_f-u_0)s^3(10-15s+6s^2),
\label{us}
\eeq
where $s=t/t_f$. See Fig. \ref{fig_0} for examples of this function and the corresponding $\omega(t)^2$.     
$u(t)$ in Eq. (\ref{us}) goes from $u_0$ to $u_f$ monotonously and posseses the symmetry $u(t_f/2+\tau)+u(t_f/2-\tau)=u_f+u_0$.  
 
The following first order moments are calculated from Eq. (\ref{wave}) by using triple 
integrals and delta-function derivatives. 
Since $u(t)$ appears only in the form of the ratio $U_t=u_t/u_0$ we can work out all expressions in terms of $U_t$, 
\beq
\hspace*{-.2cm}\la q\ra_t=\la q\ra_0 {U_t}+\la p\ra_0\frac{U_t}{m}{\cal I}_t,\;\;  
\la p\ra_t=\la q\ra_0{m\dot{U}_t}+\la p\ra_0 \frac{{\cal A}_t}{U_t}, 
\eeq
{where} ${\cal A}_t= 1+U_t\dot{U}_t{\cal I}_t$.
%
%
%
\begin{figure}
\begin{center}
\hspace*{-.2cm}\includegraphics[width=1.05\linewidth]{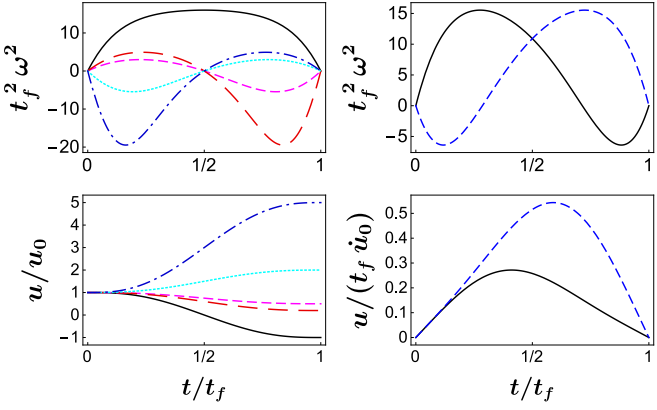}
\end{center}
\caption{(Color online) $\omega^2(t)$ (top) and corresponding $u(t)$ (bottom) for different processes:  Left: Momentum scaling; Right: Position scaling.   
Momentum scaling is designed with Eq. (\ref{us}), and scaling factors $u_0/u_f=5$ (long dashed red), $u_0/u_f=2$ (dashed magenta), $u_0/u_f=1/2$ (dotted cyan), $u_0/u_f=1/5$ (dotted-dashed blue) and $u_0/u_f=-1$ (solid black, a momentum reversing process).
Spatial scaling uses a different polynomial for $u(t)$, to satisfy: $u(t_b)=\ddot{u}(t_b)=0$, $\dot{u}(0)=\dot{u}_0$ and $\dot{u}(t_f)=\dot{u}_f$. We depict a focusing protocol, $\dot{u}_0/\dot{u}_f=-1/2$ (dashed blue), and a spreading protocol,  $\dot{u}_0/\dot{u}_f=-2$ (solid black). $\omega(t_b)=0$ in all cases.\label{fig_0}}
\end{figure}
%

\begin{figure}
\begin{center}
\hspace*{-0.5cm}\includegraphics[width=1.1\linewidth]{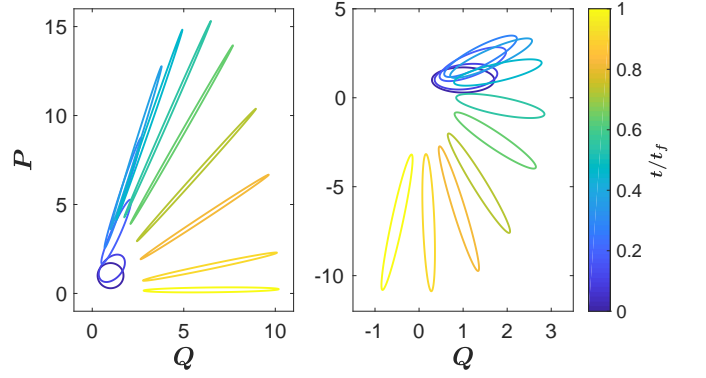}
\end{center}
\caption{(Color online) Evolution of a Gaussian state from $t=0$ to $t_f$ in phase space. Twelve snapshots at equal time intervals
of a Wigner-function contour line. The color sidebar helps to follow the time ordering from  $t=0$ (purple) to $t_f$ (yellow).   
The dimensionless units are explained in the main text. 
The initial ``off-center'' state is a minimum-uncertainty-product state. In the initial state the principal semiaxes are 
$\Delta Q=\Delta P=2^{-1/2}$. $\la P\ra_0=\la Q\ra_0=1$.  Left: Momentum scaling, 
$u_0/u_f=1/5$.  Right: Position scaling, $\dot{u}_0/\dot{u}_f=-1/2$. See corresponding $\omega^2(t)$ 
in Fig. \ref{fig_0}. 
\label{fig_1}}
\end{figure}
%
Similarly, the second order moments are 
\beqa
&&\hspace*{-1.2cm}\la q^2\ra_t=\la p^2\ra_{0}\left(\!\frac{U_t{\cal I}_t}{m}\!\right)^{\!\!2}\!\!+\la qp+pq\ra_{0}\frac{U_t^2{\cal I}_t}{m}
+\la q^2\ra_{0}{U_t^2},
\\
&&\hspace*{-1.2cm}\la pq+qp\ra_t=\la pq+qp\ra_0(1+2\dot{U}_tU_t{\cal I}_t)
\nonumber\\
&+&\la p^2\ra_0 \frac{2{\cal I}_t}{m}
{\cal A}_t +\!\la q^2\ra_0 {2mU_t\dot{U}_t},
\\
&&\hspace*{-1.2cm}\la p^2\ra_t=\la p^2\ra_0\frac{1}{U_t^2}{\cal A}_t^2 
\nonumber\\
&+&\la pq+qp\ra_0\,\frac{m\dot{U}_t}{U_t} {\cal A}_t 
+\la q^2\ra_0 \left(\!{m\dot{U}_t}\!\right)^{\!\!2}.
\eeqa 
The above first and second order moments are consistent with the invariants $G$ and $I$, 
in the sense that the expectation values of $G$ and $I$ are indeed constant with them.   
The variances for position and momentum take the form 
\beqa
(\Delta q)_t^2&=&(\Delta p)_0^2 \left(\frac{U_t{\cal I}_t}{m}\right)^{\!\!2}+(\Delta q)_0^2{U_t^2}
\nonumber\\
&+&\left(\la qp+pq \ra_0 -2 \la q\ra_0 \la p\ra_0 \right) \frac{U_t^2{\cal I}_t}{m},
\\
(\Delta p)_t^2&=&(\Delta p)_0^2 \frac{1}{U_t^2}{\cal A}_t^2+(\Delta q)_0^2\left({m \dot{U}_t}\right)^{\!2}
\nonumber\\
&+&\left(\la qp+pq \ra_0 -2 \la q\ra_0 \la p\ra_0 \right)\frac{m\dot{U}_t{\cal A}_t}{U_t}.
\eeqa
Considering that ${\cal A}_f=1$, we get at $t_f$ that  
$(\Delta p)_f^2=(\Delta p)_0^2 /U_f^2$ for any state. 
Moreover ${\cal I}_f=t_f\int_0^1 ds/ \tilde{U}(s)^2\sim t_f$, where $\tilde{U}(s)=U(t=st_f)$. 
For a packet without initial position-momentum correlations 
$(\Delta q)_f^2=  (\Delta q)_0^2 {U_f^2}+{\cal O} (t_f^2)$, in other words, a very fast process in which the $t_f^2$
term is neglected  
performs the momentum scaling preserving the uncertainty product $\Delta p_f \Delta q_f\approx \Delta p_0 \Delta q_0$. 
This comes at a price, as the maximal transient value of $|\omega^2|$ (and thus of the absolute value of the potential energy) scales as $\sim t_f^{-2}$ for short times. In other words, demanding shorter and shorter process times requires the ability to implement 
the harmonic oscillator potential for energies growing as $t_f^{-2}$. The practical limitations of the opposite, large time limit 
are due to the the first term in $\la q^2\ra_t$, which grows as $t_f^2$. Thus, large process times  need a potential implemented over a 
large spatial range. Similar limitations concern the first moments, in particular $\la q\ra_t$ should not exceed
the region where the potential may be implemented.   
In a realistic setting the harmonic potential will be realized within a temporal, spatial and energetic domain, which will determine the range of values allowed for the initial (first or second) moments so that the final and/or transient moments do not exceed the 
set limits. 

``Cooling'',  conserving  phase-space volume,  is an obvious application of the above by setting a large factor $U_f$. 
Notice that some of the constraints  of delta-kick cooling do not apply here, specifically, in DKC \cite{Chu1986,Ammann1997,Myrskog2000,Kovachy2015} the 
initial state must be centered at the origin in phase space, so that a free expansion elongates the state along a given well defined 
angle (phase line) and a  transient harmonic trap rotates the state to the horizontal (position) axis.  The present method, instead, does
not require any condition for the initial state, other than those imposed by the geometry of the actual setting and technical limits to  implement the harmonic potential. Figure \ref{fig_1} (left) shows the evolution of a state in phase space, initially a minimum uncertainty product state 
which is initially ``off center''.  
In the simulations and figures  
we use dimensionless variables for coordinates, times, or momenta,  defined from dimensional ones  
as  ${Q}={q}/l$, ${s}={t}/t_f$, ${P}={p}l/\hbar$, where $l=(\hbar t_f/m)^{1/2}$. The Schr\"odinger equation 
becomes $i\partial \Psi(Q,s)/\partial s=[{{P}^2}/{2}+{\Omega(s)^2}Q^2/{2}]\Psi(Q,s)$ where $\Omega(s)=t_f\omega(t)$,
$\Psi(Q,s)=l^{-1/2}\psi(q,t)$ and ${P}=-i\partial /\partial Q$.    

{\it Position focusing, momentum mirrors, and more.}
We may consider as well negative scaling factors with a $u(t)$ designed to avoid singularities in $\omega(t)$. 
The simplest case is $U_f=-1$ which inverts all momenta regardless of their initial sign 
and the initial state. The $u(t)$ function in Eq. (\ref{us}) is valid for this purpose as the zero of $u$ at $t_f/2$ is canceled by a 
zero of $\ddot{u}(t_f/2)$, see Fig. \ref{fig_0}.    
  
Zeros of $u(t)$ at some intermediate time $t_0>0$ might seem to imply singularities in the wave function $\psi(q,t)$ even if 
$\omega^2(t)$ remains finite. A detailed analysis though shows that cancellations occur, e.g. due to the asymptotic property
$\lim_{t\to t_0} u(t){\cal I}_t=-u_0^2/\dot{u}_{t_0}$, 
so that the singularities are in fact avoided. A simple example is a Gaussian state for which the momentum integral 
in Eq. (\ref{wave}) can be done formally.  

A second extension of the current methodology  is ``position focusing'' or antifocusing, 
namely to scale positions rather than momenta. Formally the procedure is very similar,  
with a  different design for $u(t)$ so that ${u}(t_b)=0$. Thus, the linear invariant 
(\ref{invaG}) is at initial and final times proportional to $q$. The process scaling is of the form $q_f= q_0 \dot{u}_0/\dot{u}_f$.  
In parallel with Eqs. (\ref{eigen},\ref{wave}) we work out the eigenvectors of $G(t)$ in momentum representation 
with eigenvalues $-m \dot{u}_0 q_0$, 
\beq
\phi_{q_0}(p,t)={\left(\frac{\dot{u}_0}{h\dot{u}_t}\right)^{\!\!1/2}e^{\frac{-i}{m\dot{u}_t\hbar}(mq_0\dot{u}_0+u_tp^2/2)}}
e^{-\frac{imq_0^2{\cal J}_t}{2\hbar}},
\eeq
{where} ${\cal J}_t=\int_0^t dt' {\omega_{t'}^2\dot{u}_0^2}/{\dot{u}_{t'}^2}$,
%
%
and a corresponding representation for arbitrary wave functions, $\psi(p,t)=\int dq_0 \phi_{q_0}(p,t)\la q_0|\psi(0)\ra.$
The invariant eigenvector and solution of the Schr\"odinger equation $\phi_{q_0}(t)$ evolves from an eigenvector of position, $\phi_{q_0}(p,0)=\la p|q_0\ra$, to a scaled 
version $\phi_{q_0}(p,t_f)=\left({\dot{u}_0}/{\dot{u}_t}\right)^{1/2}  \exp[-{imq_0^2{\cal J}_f}/({2\hbar})]   \la p|q_0\dot{u}_0/\dot{u}_t\ra$.
{For completness, the first moments are 
$\langle p\ra_t=(-m{\cal J}_t \la q\ra_0+ \la p\ra_0)({\dot{u}_t}/{\dot{u}_0})$, 
and $\la q\ra_t=\la p\ra_0 u_t/(m\dot{u}_0)+\la q\ra_0[\dot{u}_0/\dot{u}_t-(u_t/\dot{u}_0){\cal J}_t]$}.
These processes may lead to position focusing or to position expansions that can be combined with side inversions
if the scaling factor $\dot{u}_0/\dot{u}_f$ is made negative, see Fig. \ref{fig_0}.  
Again, the initial state is arbitrary.  A process for focusing  with side inversion is depicted in Fig. \ref{fig_1} 
for an initially off-center state.

So far we have considered, in all examples and boundary conditions, processes  from 
free motion to free motion, i.e., $\omega(t_b)=0$. However the frequencies at the boundaries may have any 
desired value by choosing $u(t_b)$ and its derivatives consistently. 
Specifically for momentum scaling,  $G(t_b)=u(t_b)p$ is valid as long as $\dot{u}(t_b)=0$, so 
$\omega(t_b)=0$ is not necessary. Thus  the approach can be adapted 
to scale the momenta from a trap with $\omega_0=\omega(0)$ to a trap with $\omega_f=\omega(t_t)$. 
Also the kinetic energy is scaled but not necessarily the total energy. A possible application could be 
to control the temperature  if its final desired  value   does not correspond to the one of an adiabatic
process.      
The momentum does not commute with $H(t_b)$ 
for nonzero $\omega(t_b)$ so the final momenta will not be conserved for $t>t_f$ unless the trap is switched off
abruptly at $t_f$.  As for position scaling, its combination with a nonzero $\omega(t_b)=0$ provides a way to scale the potential energy at will, since the quadratic invariant $I$ becomes proportional to the potential energy at boundary times in these protocols.  

{\it Discussion.}
Spreads of momentum or velocity of initial particles often lead to particle loss and inefficiencies 
in focusing, slowing or acceleration processes. 
Shortcuts to adiabaticity techniques can be made very robust with respect to initial conditions or protocol imperfections.
This feature and the possibility to choose and shorten the process time make them  powerful tools to design cooling \cite{Chen2010_063002,Sagesser2020,Bertolotta2020}, even for open systems \cite{Martinez2016,Villazon2019,Dann2019,Alipour2019}, 
launching  \cite{Tobalina2017}, or compression and expansion protocols \cite{Torrontegui2013,Guery2019}. 
This work, in particular, demonstrates that, making use of linear invariants, 
momentum or position scaling, irrespective of initial conditions of the particle, can be achieved. 
The proposed methodology can be adapted to   
sequential interactions for beam control,  or for trapped particles, 
for example providing a robust alternative to DKC 
to reach picokelvin temperatures.
\acknowledgments{This work was supported by the Basque Country Government (Grant No. IT986-16), and 
by  Grants  PGC2018-101355-B-I00 (MCIU/AEI/FEDER,UE) and FIS2016-80681P.}
\bibliography{Bibliography}{}
\bibliographystyle{sofia} 
\end{document}